\documentclass[twocolumn,prd,nofootinbib,showpacs,preprintnumbers]{revtex4}
\usepackage{epsfig}

\def\barp{\bar{p}}

\def\d{{\rm d}}
\def\Emax{E_{\rm max}}
\def\cA{\mathcal{A}}
\def\cB{\mathcal{B}}
\def\cE{\mathcal{E}}
\def\cF{\mathcal{F}}
\def\cO{\mathcal{O}}

\begin{document}
\title{High-energy antiprotons from old supernova remnants}
\author{Pasquale Blasi$\,{}^{p}$ and Pasquale D.~Serpico$\,{}^{\bar p}$}
\affiliation{$^{p}$INAF-Osservatorio Astrofisico di Arcetri, Largo E. Fermi, 5 50125 Firenze, Italy}
\affiliation{$^{\bar p}$Physics Department, Theory Division,
CERN, CH-1211 Geneva 23, Switzerland}

\date{\today}

\begin{abstract}
A recently proposed model~\cite{Blasi:2009hv} explains the rise in energy of the positron fraction measured by the PAMELA satellite in terms of hadronic production of positrons in aged supernova remnants, and acceleration therein. Here we present a preliminary calculation of the anti-proton flux produced by the same mechanism. While the model is consistent with present data, a rise of the antiproton to proton ratio is predicted at high energy, which strikingly distinguishes this scenario from other astrophysical explanations of the positron fraction (like pulsars). We briefly discuss important implications for Dark Matter searches via antimatter.
\end{abstract}

\maketitle
{\it Introduction}---The antimatter component in cosmic rays (CRs) has been recognized since long time as an important diagnostic tool for cosmology (e.g. matter-antimatter asymmetry of the local universe), particle physics (indirect dark matter searches), and properties of cosmic
ray sources and propagation medium (see the standard textbooks~\cite{Ginzburg:1990sk} or the more recent reviews~\cite{Maurin:2002ua,Strong:2007nh}). Recently, the PAMELA satellite detector~\cite{pamela} has presented its first results of the measurement of the positron fraction in the cosmic ray spectrum, which appears to begin climbing quite rapidly between $\sim 7$ GeV and  100 GeV~\cite{Adriani:2008zr}. This trend confirms (with much higher statistics and over a wider energy range) what previously found by other experiments, including HEAT~\cite{heat} and AMS-01~\cite{ams01}.  On very general grounds, this behavior is at odds with the standard predictions for secondary positrons produced in the collisions of cosmic ray nuclides with the inter-stellar medium (ISM);  an additional source of positrons seems to be required~\cite{Serpico:2008te}. While numerous models of dark-matter (DM) annihilation or decay have been proposed (for a complete list see refs. to~\cite{pamela}), astrophysical explanations do exist, in particular invoking $e^{+}-e^{-}$ acceleration in pulsars~\cite{pulsars}.

A generic feature of these astrophysical solutions is the absence of a significant anti-proton signal accompanying the positron one, since the acceleration involves purely electromagnetic phenomena in pulsar magnetospheres. Actually, the PAMELA collaboration has presented also data on the $\barp/p$ ratio below $E\sim 100\,$GeV~\cite{Adriani:2008zq}, which fit naturally in a scenario of purely secondary production via CR spallation in the interstellar medium (ISM). In turn, this  puts a non-trivial constraint on dark matter models trying to account for the positron excess~\cite{Donato:2008jk}.

Recently, one of us has proposed an alternative and even simpler astrophysical explanation for the feature observed in the positron fraction~\cite{Blasi:2009hv}. In this scenario, the `excess' is due to positrons created as secondary products of hadronic interactions inside the standard sources of CRs, supernova remnants (SNRs). In particular, positrons would be produced in the late stage of SNR evolution, when also the bulk of cosmic rays (namely below the knee) are expected to be accelerated. The crucial physical ingredient which leads to a natural explanation of the positron flux is the fact that the secondary production takes place in the same region where cosmic rays are being accelerated; secondary $e^{+}$ (and $e^{-}$) participate in the acceleration process and turn out to have a very flat spectrum at high energy, which is responsible, after propagation in the Galaxy, for the observed positron ``excess''. The values of the parameters which lead to an explanation of the rising positron fraction are typical of old SNRs, rather than the young, often gamma-ray and X-ray bright ones. Since this is now a {\it hadronic} mechanism for the explanation of the data, one expects an associated feature in the antiproton spectrum. The purpose of this letter is to present a preliminary calculation of the $\barp/p$ ratio within the simple model of~\cite{Blasi:2009hv}. It is important to realize that this model applies to a stage of the SNR evolution in which: 1) not many observations are available, with the possible exception of the ones in the radio band, 2) many effects are expected to play a role, such as magnetic field damping, on which we have exceedingly poor control, and 3) it would be important to carry out the calculations in a time dependent way, in order to move beyond a simple estimate. Thus, some of the
parameters adopted in~\cite{Blasi:2009hv} might be considered as ``effective'' astrophysical inputs. While they need to be checked versus more realistic models, for the time being we believe that a more urgent task is to establish whether the model of~\cite{Blasi:2009hv} can account for the rising trend of the PAMELA data; this can be done most reliably  by checking it versus independent predictions of the same model. The calculation of the antiproton flux (or, rather, the antiproton to proton ratio) is the first of them and the one requiring the minimum number of independent assumptions. Such a calculation reveals that: i) the additional signal does not violate existing data; ii) a {\it generic} prediction of the model is a flattening and eventually a weak rise of the $\barp/p$ ratio in the decade $\sim 100\div 1000$ GeV; the value of this ratio at TeV energy is about one order of magnitude above expectations from the conventional models (of course, this is equivalent to saying that a spectral break is predicted in the absolute $\barp$ spectrum). Since this feature is strictly related to the one in the positron spectrum, the model is very predictive. Clearly, these results have very important implications also for dark matter searches via anti-protons as well as for astrophysical diagnostics via the $\barp/p$ ratio, as we shall comment at the end of the letter.

{\it The calculation}---Here we only report the specific equations that are needed to calculate the $\barp/p$  spectrum, while referring to~\cite{Blasi:2009hv} for a detailed description of the model.  It is also worth stressing that up to several tens of GeV the  $\barp/p$  spectrum is well in agreement with conventional mechanisms~\cite{Adriani:2008zq}. Since we are only interested in the high energy part of the spectrum, $E\gg 10\,$GeV, we shall implicitly assume approximations valid in the relativistic limit. Compared with the treatment for positron production proposed in ref.~\cite{Blasi:2009hv}, the differences arise only in the production cross-section and in the physical processes relevant for propagation. Concerning propagation, for the purposes of this letter we can neglect energy losses, re-acceleration, tertiary production, solar modulation, etc. which are relevant at low energy. The antiprotons are injected inside the sources as described by the function
\begin{equation}
Q_{\barp}(E) \simeq 2 \int_{E}^{E_{\rm max}} \d\cE  N_{CR}(\cE) \sigma_{p\barp}(\cE,E) n_{gas}\, c\,,\label{master1}
\end{equation}
where $c$ is the speed of light, $n_{gas}$ is the gas density for $pp$ scattering in the shock region and $\sigma_{p\barp}(\cE,E)$ is the differential cross section for a proton of energy $\cE$ to produce a $\bar{p}$ of energy $E$. The energy $E_{\rm max}$ is the maximum energy of the protons being accelerated in the SNR at the age relevant for the mechanism discussed here and it is discussed in the following. The factor 2 accounts for the antiproton coming from antineutron production, which we assume to be identical to the $\barp$  one (isospin symmetry limit).
A subtle point is that antineutrons, being neutral, stream freely away from the acceleration region until they decay (barring nuclear collisions). The range of an (anti-)neutron of energy $E$ is $R_n\simeq (E/m_n\,c^2)\tau_n\,c\simeq 10^{-5}\,E_{\rm GeV}\,$pc, where $\tau_n$ and $m_n$ are the lifetime and mass of
the neutron, respectively. The following considerations assume that the confinement/acceleration region has a characteristic size $\ell\gg R_n$, which is easily fulfilled for pc-scale shocks in SNRs (see also below).
As in most calculations in the modern literature, for the cross section $\sigma_{\barp}(\cE,E)$ we use the parameterization of Ref.~\cite{Tan:1982nc}. 
After production, the spectrum described by Eq.~(\ref{master1}) is modified by acceleration inside the source and by propagation to the Earth. The latter phase is identical for $\barp$ and $p$, so the spectral modification induced by propagation cancels in the antiproton to proton ratio. We assume that SNRs account for the overall CR flux at the Earth and, for the moment, we are assuming that it is entirely made of protons. Also, throughout the paper we are relying on the fact that most of the GeV-TeV production of cosmic ray protons happens in the late stage of SNRs which is of concern here. Then, the solution for  the $\barp/p$ flux ratio can be easily derived from~\cite{Blasi:2009hv} in the form
\begin{equation}
\frac{J_{\barp,SNRs}(E)}{J_{p}(E)} \simeq 2\,n_{1}\,c\,[\mathcal{A}(E)+\mathcal{B}(E)]\label{master2}
\end{equation}
where
\begin{eqnarray} 
&&\mathcal{A}(E)=\gamma\left(\frac{1}{\xi}+r^2\right)\times \\
&&\times\int_m^{E}\d\omega \,{\omega}^{\gamma-3} \frac{D_{1}(\omega)}{u_1^2}\int_{\omega}^{E_{\rm max}} \d \cE\,{\cE}^{2-\gamma}\,\sigma_{p\barp}(\cE,\omega) \,,\label{SNRpbarA}
\end{eqnarray}
and
\begin{equation} 
\mathcal{B}(E)=\frac{\tau_{SN}\,r}{2\,E^{2-\gamma}}\int_{E}^{E_{\rm max}}  \d \cE\,{\cE}^{2-\gamma}\,\sigma_{p\barp}(\cE,E)\,\label{SNRpbarB}.
\end{equation}
In the above expressions,  $n_{1}$ is the background gas target in the upstream region of the shock, $u_1$ the fluid velocity there (which we fix at $u_1=0.5\times 10^{8}\,$cm/s), $\tau_{SN}$ is a typical SNR age, here fixed to $\tau_{SN}=2\times 10^4\,$yr.  The parameter $\xi$ (which we fix as $\xi\simeq 0.17$) is the fraction of proton energy carried away by a secondary antiproton, while $r$ is the compression factor between upstream and downstream.
The index  $-\gamma$ is the slope of the spectrum in {\it momentum} space, which is related to the spectral index $\alpha$ in energy space of the accelerated cosmic ray protons at the source via $\alpha=2-\gamma$ and to the ratio $r$ by $\gamma=3\,r/(r-1)$. Here, we fix $r=3.8$ so that $\alpha\simeq -2.07$.  Note that this is another instance of the oversimplification we are forced to here: the compression factor $r$ is chosen in order to achieve an injection spectrum $\propto E^{-2.1}$ necessary to fit the CR spectrum after propagation. However, it is well known that SNR shocks stay strong ($r=4$) at almost all times. It follows that the spectrum steeper than $E^{-2}$ should follow from a complex overlap over time during the SNR evolution rather than the fact that the shock is weaker. If taken into account, this effect leads to an enhanced rate of production of secondaries inside the SNR for a given set of parameters. 

 Finally, the function $D_{1}(\omega)$ is the diffusion coefficient upstream of the shock, which in quasi-linear theory writes
\begin{equation} 
D_1(E)=\left(\frac{\lambda_c\,c}{3\,\mathcal{F}}\right)\left(\frac{E}{e\,B\,\lambda_c}\right)^{2-\beta}\,,
\end{equation}
where $B$ is the magnetic field, $e$ the unit charge, $\cF\sim (\Delta B/B)^2$ is the ratio of power in turbulent magnetic field over that in the ordered one, $\lambda_{c}$ is the largest coherence scale of the turbulent component, and $\beta$ is the index carachterizing the fluctuation spectrum. For simplicity, in the following we fix $\beta=1$ (Bohm-like diffusion index), in which case the model does not depend explicitly on $\lambda_{c}$ (for a Kraichnan spectrum, one would have $\beta=1/2$
and a dependence from the square root of both $E$ and $\lambda_c$). Denoting by $B_{\mu {\rm G}}$ the magnetic field in micro-Gauss and by $E_{\rm GeV}$ the energy in GeV, numerically one has
\begin{equation} 
D_1(E)\simeq 3.3\times 10^{22}\mathcal{F}^{-1} E_{\rm GeV} B_{\mu {\rm G}}^{-1}\,{\rm cm}^2\,{\rm s}^{-1}\,.\label{bohm}
\end{equation}

For the following numerical estimate, we fix $n_{1}=2$ (in cm$^{-3}$) and  $\{\cF,B_{\mu {\rm G}}\}=\{1/20,1\}$. We consider these numbers as reasonable if applied to old SNRs, in which magnetic field amplification is not effective and in fact it is likely that magnetic fields are damped (see for instance \cite{zira2005}). We stress once again that this period is very poorly modeled and a precise quantification of the astrophysical parameters is tricky: for instance damping is required to lower the maximum energy of accelerated particles, but the temporal dependence of the maximum energy is not known, though it is expected to be rather fast. The velocity of the shock $u_1$ is better known, since it can be estimated by using the standard Sedov solution in a constant density of the background medium, yet the new term is quite sensitive to it (depending on $u_1^2$). More complicated situations---such as the expansion in a density profile induced by a presupernova wind---are of relevance only in the early stages of the expansion of the shell, and in any case only for supernovae of type II. All in all, we are using simple effective parameters with all the limitations that this approach implies. More important for the phenomenology is that the combination of parameters $n_{1}B_{\mu {\rm G}}^{-1}u_{8}^{-2}/\cF\approx 160$ is roughly what required to fit the high-energy behavior of the 
positron fraction, within a fudge factor of $\cO(1)$.
Note that, for the chosen parameters and the energy range we are interested in, the characteristic size of acceleration $\ell\simeq D_1/u_1$ is roughly three orders of magnitude larger than $R_n$, confirming a posteriori the validity of including  $\bar{n}$ in the source term. 
Another important point to discuss is that of the maximum energy for the primary and secondary particles: protons accelerated at the shock have a maximum energy which in principle can be estimated by equating the acceleration time and the age of the remnant. The maximum energy of secondary products is determined by the process responsible for their production: for electrons and positrons, typically the energy of the secondaries is $\xi\sim 0.05$ of the parent proton. For antiprotons this fraction is $\xi\sim 0.17$. However those secondary particles which are produced within a distance of order $D(E)/u$ on both sides of the shock participate in the acceleration process and they end up being accelerated at roughly the same maximum energy as the parent protons, with a rather flat spectrum. For typical values of parameters one can easily find maximum energy in the range between 3 TeV (for Bohm) and 80 TeV (for Kraichnan). However these numbers do not take into account a number of phenomena, such as the presence of higher energy particles generated at previous times, damping of the field and the possibility of obliquity of the magnetic field lines over most of the shock surface. Because of these numerous uncertainties, we adopt a sort of effective value of $10$ TeV for $\Emax$, though one has to keep in mind that all limitations listed above.  As long as $E_{\rm max}=\cO(10)\,$TeV (within a factor a few), its exact value is of minor impact for predictions of $\bar{p}/p$ at $E\alt 1\,$TeV.
Qualitatively, a higher value of $\Emax$ would increase the slope of the rise in the positron ratio, while a lower value would flatten it.

Finally, we comment on the role of nuclei in our calculations: in~\cite{Gaisser:1991vn} it was found that for a typical composition mixture like the one measured locally, correcting for this effect roughly amounts to a factor $\varepsilon\simeq 1.20$. We have  repeated the effective weighted-average renormalization using the same cross-section weights as in~\cite{Gaisser:1991vn}, but the updated composition ratio compiled in~\cite{pdg}, table 24.1. We obtain a factor $\varepsilon=$1.26, which we shall use to renormalize both Eq.~(\ref{master2}) and the ISM contribution. The ISM spallation contribution to $\barp/p$ can be written in the same approximation as above in the form
\begin{equation}
\frac{J_{\barp,ISM}(E)}{J_{p}(E)} \simeq \frac{2\,\varepsilon\,X(E)}{m_p\,{E}^{2-\gamma-\delta}} \int_E^\infty \d\cE {\cE}^{2-\gamma-\delta} \sigma_{p\barp}(\cE,E) \,,\label{ISMsec}
\end{equation}
 with the grammage parameterized as: 
\begin{equation}
X(E)=\Gamma\left(\frac{E}{10\,{\rm GeV}}\right)^{-\delta}{\rm g}\,{\rm cm}^{-2}\:\:\:\:(E\geq 10\,{\rm GeV})\,.
\end{equation}
Here we adopt $\delta=0.6$ and $\Gamma=5.5$, well within the range discussed in~\cite{GarciaMunoz:1987}.
\begin{figure}[!thb]
\begin{center}
\begin{tabular}{c}
\epsfig{figure=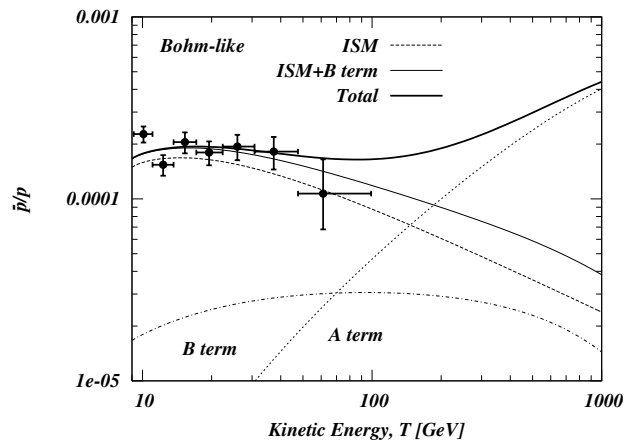,width=0.98\columnwidth}
\end{tabular}
\end{center}
\vspace{-0.9pc} \caption{
The $\barp /p$ ratio for the parameters reported in the text, together with a simple model of secondary production in the ISM (dashed line), and with the recent data from PAMELA~\cite{Adriani:2008zq}.  The dotted and dot-dashed lines represent the contributions of the $\cA$ and $\cB$ terms of Eq.~(\ref{master2}) alone, respectively. The thick solid curve is the overall contribution due to ISM plus the new mechanism, while the thin solid curve only includes the ISM contribution plus the $\cB$ term.}\label{results}
\end{figure}

The predictions thus obtained are reported in Fig.~\ref{results}. Although the model described is very simple, the overall agreement with the data is good, with the predictions for the conventional model (only antiprotons from spallation in the ISM) strongly differing from the present ones beyond the $E\sim$100 GeV region. In the case considered here the ratio flattens at first, then grows with energy. The latter behavior is exclusively due to the~$\cA$-term of Eq.~(\ref{SNRpbarA}), while the former behavior is due to the interplay of the decreasing ISM term of Eq.~(\ref{ISMsec}), the rising $\cA$-term and the relatively flat $\cB$-term of~Eq.~(\ref{SNRpbarB}). It is important to note that the $\cB$-term accounts for production of $\barp$ without ``reacceleration'', thus it does not depend on the diffusion properties, only on the density of the environment $n_1$ and the typical timescale $\tau_{SN}$: we see that its presence alone contributes to change appreciably the shape of $\barp/p$ at  $E\agt$100 GeV. The role of the $\cA$-term is even
more dramatic at high-energy (and indeed it is essential to explain a positron fraction rise of the kind revealed by PAMELA), but it is somewhat more model dependent both in shape and normalization.

{\it Discussion and Conclusions}---In this letter, we have discussed an important signature of the  mechanism proposed in~\cite{Blasi:2009hv} to explain the anomalous behavior of the positron ratio at high energies: a harder component should emerge in the antiproton spectrum at energies above $\sim 100$ GeV. The $\bar p/p$ ratio flattens at first, then eventually starts
rising with energy. New data at high energy from PAMELA and AMS-02~\cite{AMS-02} should easily distinguish between this explanation and a pulsar related one for the positron fraction. 
As discussed above, though the effect predicted here (and in \cite{Blasi:2009hv} for positrons) must be present, its strength depends on the many parameters of the problem and on whether they are appropriate to describe the final stages of SNRs. This uncertainty is mainly of concern for the rising (``reacceleration'') term, while the injection term is less model dependent. The latter shows-up as a flattening in the antiproton ratio and represents a conservative prediction for the energies just above the ones currently probed by PAMELA. Even limiting ourselves to the effects of the latter term, the implications for astrophysics are of crucial
importance: The good news is that the high-energy range of the antiproton spectrum may reveal important constraints on the physics of the CR acceleration sites. The bad news is that it is not straightforward to infer from high energy  $\bar p/p$-data the propagation parameters, as the diffusion index $\delta$, since they are partially degenerate with source parameters: The thin solid line in Fig.~\ref{results} might be easily confused with a purely ISM model with no contribution at all from SNRs, but a lower value of the diffusion index in the ISM, $\delta$.

Similarly, our results may change dramatically the perspectives for the {\it detection} of DM via a signature in high-energy antiprotons: 
Indeed, we have discussed a purely astrophysical mechanism to produce a high-energy ``excess'' of antiprotons over the secondary yield from ISM production. 
Even a subleading role for the mechanism proposed in~\cite{Blasi:2009hv} in explaining the positron excess might produce measurable anomalies in the antiproton 
spectrum.  An ``excess'' in the high-energy range of  $\bar p/p$ could not be interpreted anymore uniquely as manifestation of new physics: compare for example Fig.~3 in~\cite{Donato:2008jk} with our Fig.~\ref{results}. The mechanism proposed here might thus require a paradigm change for DM searches via anti-matter, at least until the contribution from standard astrophysical sources is understood and corresponding uncertainties are kept under control.



\begin{thebibliography}{00}

\bibitem{Blasi:2009hv}
  P.~Blasi,
  ``The origin of the positron excess in cosmic rays,''
  arXiv:0903.2794.

 
    \bibitem{Ginzburg:1990sk}
V.~L.~Ginzburg {\it et al.},
  ``{\it Astrophysics of cosmic rays},''
{\it  Amsterdam, Netherlands: North-Holland (1990)};
  T.~K.~Gaisser,
``{\it Cosmic rays and particle physics},''
{\it  Cambridge, UK: Univ. Pr. (1990) 279 p}


\bibitem{Maurin:2002ua}
  D.~Maurin {\it et al.}, 
  ``Galactic cosmic ray nuclei as a tool for astroparticle physics,''
  in Research Signposts, "{\it Recent Research Developments in Astrophysics}" [astro-ph/0212111]
\bibitem{Strong:2007nh}
  A.~W.~Strong, I.~V.~Moskalenko and V.~S.~Ptuskin,
  ``Cosmic-ray propagation and interactions in the Galaxy,''
  Ann.\ Rev.\ Nucl.\ Part.\ Sci.\  {\bf 57}, 285 (2007).
 
\bibitem{pamela}
\url{http://pamela.roma2.infn.it/index.php}

\bibitem{Adriani:2008zr}
  O.~Adriani {\it et al.},
  ``Observation of an anomalous positron abundance in the cosmic radiation,''
 {\it Nature}  {\bf 458}, 607 (2009). 

\bibitem{heat}
  J.~J.~Beatty {\it et al.},
  ``New measurement of the cosmic-ray positron fraction from 5-GeV to 15-GeV,''
  Phys.\ Rev.\ Lett.\  {\bf 93}, 241102 (2004).

\bibitem{ams01}
  M.~Aguilar {\it et al.},  
  ``Cosmic-ray positron fraction measurement from 1-GeV to 30-GeV with AMS-01,''
  Phys.\ Lett.\  B {\bf 646}, 145 (2007).
  

\bibitem{Serpico:2008te}
  P.~D.~Serpico,
  ``Possible causes of a rise with energy of the cosmic ray positron fraction,''
  Phys.\ Rev.\  D {\bf 79}, 021302 (2009).

\bibitem{pulsars}
  D.~Hooper, P.~Blasi and P.~D.~Serpico,
  ``Pulsars as the Sources of High Energy Cosmic Ray Positrons,''
  JCAP {\bf 0901}, 025 (2009);
  H.~Yuksel, M.~D.~Kistler and T.~Stanev,
  ``TeV Gamma Rays from Geminga and the Origin of the GeV Positron Excess,''
  arXiv:0810.2784;
  S.~Profumo,
  ``Dissecting Pamela (and ATIC) with Occam's Razor [\ldots]''
  arXiv:0812.4457;
  D.~Malyshev, I.~Cholis and J.~Gelfand,
  ``Pulsars versus Dark Matter Interpretation of ATIC/PAMELA,''
  arXiv:0903.1310.
  
\bibitem{Adriani:2008zq}
  O.~Adriani {\it et al.},
  ``A new measurement of the antiproton-to-proton flux ratio up to 100 GeV in
  the cosmic radiation,''
  Phys.\ Rev.\ Lett.\  {\bf 102}, 051101 (2009).
  
    \bibitem{Donato:2008jk}
  F.~Donato {\it et al.}
  ``Constraints on WIMP Dark Matter from the High Energy PAMELA $\bar{p}/p$
  data,''
  Phys.\ Rev.\ Lett.\  {\bf 102}, 071301 (2009).

\bibitem{Tan:1982nc}
  L.~C.~Tan and L.~K.~Ng,
  ``Parametrization Of Anti-P Invariant Cross-Section In P P Collisions Using A
  New Scaling Variable,''
  Phys.\ Rev.\  D {\bf 26}, 1179 (1982).
  

\bibitem{Gaisser:1973nz}
  T.~Gaisser \& R.~Maurer,
  ``Cosmic anti-p production in interstellar p p collisions,''
  Phys.\ Rev.\ Lett.\  {\bf 30}, 1264 (1973).


\bibitem{Gaisser:1991vn}
  T.~Gaisser \& R.~Schaefer,
  ``Cosmic ray secondary anti-protons: A Closer look,''
  Astrophys.\ J.\  {\bf 394}, 174 (1992).

\bibitem{pdg}
``Cosmic Rays'' review by T. Stanev and T. Gaisser in
C. Amsler {\it et al.}, Physics Letters B667, 1 (2008)  [PDG]

\bibitem{zira2005}
V.S. Ptuskin and V.N. Zirakashvili, 
`` On the spectrum of high-energy cosmic rays produced by supernova remnants in the presence of strong cosmic-ray streaming instability and wave dissipation.''
A\&A {\bf 429}, 755 (2005).


\bibitem{GarciaMunoz:1987}
  M.~Garcia-Munoz {\it et al.}, 
 ``Cosmic-ray propagation in the Galaxy and in the heliosphere - The path-length distribution at low energy'',
  Astrophys.\ J.\ Suppl. {\bf 64}, 269 (1987).


\bibitem{AMS-02}\url{http://ams.cern.ch/}

\end{thebibliography}
\end{document}